\documentclass[aps,twocolumn,floatfix,groupedaddress,nofootinbib]{revtex4}
\usepackage{csquotes}
\usepackage{graphicx,color}
\usepackage{float}
\usepackage[caption=false]{subfig}
\usepackage{amsmath}
\usepackage{amssymb}
\usepackage{multirow}
\usepackage{dsfont}
\usepackage{tikz}
\expandafter\let\csname equation*\endcsname\relax
\expandafter\let\csname endequation*\endcsname\relax

\begin{document}

\title{Fourier--Matsubara series expansion for imaginary--time correlation functions}
\author{Panagiotis Tolias$^{1}\footnotemark\footnotetext{corresponding author: tolias@kth.se}$, Fotios Kalkavouras$^{1}$ and Tobias Dornheim$^{2,3}$}
\affiliation{$^1$Space and Plasma Physics - Royal Institute of Technology (KTH), SE-10044 Stockholm, Sweden\\
             $^2$Center for Advanced Systems Understanding (CASUS), D-02826 G\"orlitz, Germany\\
             $^3$Helmholtz-Zentrum Dresden-Rossendorf (HZDR), D-01328 Dresden, Germany}
\begin{abstract}
\noindent  A Fourier--Matsubara series expansion is derived for imaginary--time correlation functions that constitutes the imaginary--time generalization of the infinite Matsubara series for equal-time correlation functions. The expansion is consistent with all known exact properties of imaginary--time correlation functions and opens up new avenues for the utilization of quantum Monte Carlo simulation data. Moreover, the expansion drastically simplifies the computation of imaginary--time density--density correlation functions with the finite temperature version of the self-consistent dielectric formalism. Its existence underscores the utility of imaginary--time as a complementary domain for many-body physics.
\end{abstract}
\maketitle

\emph{-- Motivation}. The theoretical modeling of the dynamic properties of correlated quantum many-body systems has proven to be notoriously difficult. Exact quantum Monte Carlo (QMC) simulations of finite temperature systems are typically restricted to the imaginary--time domain, thus they provide direct access to imaginary--time correlation functions (ITCFs) that correspond one-to-one to spectral functions (SFs) in the real frequency domain\,\cite{introQMC1,introQMC2,introQMC3}. The connection between ITCFs and SFs is a two-sided Laplace transform and constitutes the starting point for \emph{analytic continuation}, i.e., the numerical inversion of the QMC ITCF data to acquire the sought-for SF data. Such an SF extraction is equivalent to the numerical inversion of the Laplace transform; a well-known ill-posed problem subject to a number of numerical instabilities\,\cite{introAC01,introAC02}. In fact, considering the unavoidable QMC error bars and the necessarily finite imaginary--time resolution, the unique determination of SFs with sufficient accuracy becomes a formidable task. Although several inversion techniques have been developed, each bears its own advantages and disadvantages\,\cite{introAC03,introAC04,introAC05,introAC06,introAC07,introAC08,introAC09,introAC10}. It is worth pointing out that the availability of exact results for model systems such as the uniform electron gas (UEG) has led to successful analytical continuation based on stochastic sampling with rigorous constraints imposed on the trial solutions\,\cite{introUEG1,introUEG2}.

In the context of warm dense matter (WDM)\,\cite{introWD01,introWD02,introWD03}, it has been recently argued that, since ITCFs and SFs encode the same physics given the uniqueness of two-sided Laplace transforms, a \emph{paradigm shift} from the frequency domain of the SFs to the imaginary--time domain of the ITCFs would circumvent the challenge of analytic continuation\,\cite{introWD04,introWD05,introWD06,introWD07}. In fact, imaginary--time manifestations of several spectral signatures, such as sharp quasi-particle peaks or subtle roton features, have been identified\,\cite{introWD05}.

With respect to the \emph{experimental front}, such a domain shift has already proven to be very beneficial\,\cite{introWD08,introWD09,introWD10}. X-ray Thomson scattering (XRTS) is a widely used diagnostic of the electronic density response of WDM\,\cite{introWD11,introWD12}, where the detected beam intensity is proportional to the convolution of the dynamic structure factor (DSF), which is the SF that characterizes the electronic density, with the combined source and instrument function (SIF). In the frequency domain, the numerical deconvolution necessary for DSF extraction is unstable with respect to the experimental noise, thus a forward modelling procedure is typically employed that involves fitting of the unknown thermodynamic variables and is based on the Chihara decomposition into bound and free electrons\,\cite{introWD13,introWD14,introWD1f}. On the other hand, the shift to the imaginary--time domain allows the direct removal of the SIF effects (since convolution turns into a multiplication) and the accurate extraction of the density--density ITCF (since the two-sided Laplace transform turns out to be robust with respect to the experimental noise)\,\cite{introWD08,introWD09}. This experimental adaptation of the imaginary--time domain very recently culminated into the first model-free QMC-based interpretation of warm dense beryllium XRTS measurements\,\cite{introWD15} that were carried out at the National Ignition Facility\,\cite{introWD16}.

With respect to the \emph{theoretical front}, there exist frameworks that should translate to the imaginary--time domain in a straightforward fashion. An important example concerns the self-consistent dielectric formalism\,\cite{NoziPines,SingwTosi,IchiRepor,IchiRevMP}; a sophisticated versatile approach based on linear response theory\,\cite{quantelec} that has provided very accurate results for the thermodynamic and static properties of the finite temperature UEG in the warm dense\,\cite{TanakaIch,qSTLSsche,QMCresul1,QMCresul2,ESAschem1,ESAschem2,ESAschem3,VSscheme1,VSscheme2} and strongly coupled regimes\,\cite{HNCschem1,HNCschem2,IETschem1,IETschem2,qIETschem}. In what follows, we consider the arbitrary operator $\hat{A}$ (not necessarily Hermitian) instead of introducing $\hat{A}\equiv\hat{\rho}_{\boldsymbol{q}}$ as done in the dielectric formalism (with $\hat{\rho}_{\boldsymbol{q}}$ the microscopic one-particle density operator in Fourier space) in order to highlight the generality of the results. The finite temperature version of the dielectric formalism combines the zero frequency moment sum rule for the SF, $S_{\mathrm{AA}^{\dagger}}(\omega)=(2\pi)^{-1}\int_{-\infty}^{+\infty}\langle\hat{A}(t)\hat{A}^{\dagger}\rangle_0e^{\imath\omega{t}}$, that yields the equal-time correlation function also known as static structure factor (SSF), $S_{\mathrm{AA}^{\dagger}}=\langle\hat{A}\hat{A}^{\dagger}\rangle_0$,
\begin{equation}
S_{\mathrm{AA}^{\dagger}}=\int_{-\infty}^{+\infty}S_{\mathrm{AA}^{\dagger}}(\omega)d\omega\,,\label{zerofreq}
\end{equation}
with the quantum version of the fluctuation--dissipation theorem (FDT) that involves the linear response function $\chi_{AA^{\dagger}}(\omega)$\,\cite{quantelec}
\begin{equation}
S_{\mathrm{AA}^{\dagger}}(\omega)=-\frac{\hbar}{\pi}\frac{\Im\{\chi_{AA^{\dagger}}(\omega)\}}{1-e^{-\beta\hbar\omega}}\,,\label{qFDT}
\end{equation}
which ultimately leads to an infinite series representation of the SSF through the analytically continued linear response function $\widetilde{\chi}_{AA^{\dagger}}(z)$ evaluated at the imaginary bosonic Matsubara frequencies $\imath\omega_l=2\pi\imath{l}/\beta\hbar$\,\cite{introWD03,TanakaIch}
\begin{equation}
S_{\mathrm{AA}^{\dagger}}=-\frac{1}{\beta}\sum_{l=-\infty}^{+\infty}\widetilde{\chi}_{AA^{\dagger}}(\imath\omega_l)\,.\label{SSF_Matsubara}
\end{equation}
As it currently stands, following the evaluation of SSFs via a closed set of non-linear integral equations, the computation of density--density ITCFs in the finite temperature dielectric formalism involves a cumbersome passage from discrete imaginary frequencies to continuous real frequencies to continuous imaginary times. The intermediate step would have been redundant, had there been a link between imaginary frequencies and imaginary times.

In this work, we derive a Fourier--Matsubara series expansion for the imaginary--time correlation functions that constitutes the imaginary--time generalization of the Matsubara series for equal-time correlation functions, see Eq.(\ref{SSF_Matsubara}). We discuss its consistency with all known properties of imaginary--time correlation functions. We demonstrate how its application in the framework of the finite temperature dielectric formalism leads to the immediate evaluation of density--density ITCFs. Finally, we discuss applications concerning the utilization of QMC results.

\begin{figure}
	\centering
	\includegraphics[width=3.00in]{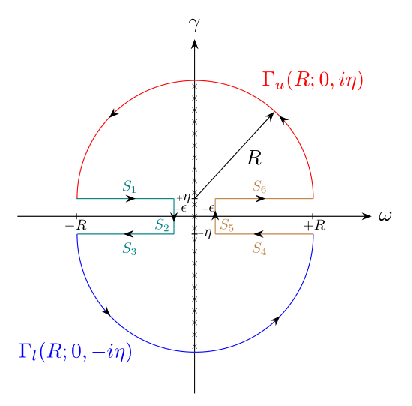}
	\caption{Illustration of the complex frequency plane parametric integration contour $\mathcal{C}(\epsilon,\eta,R)$, see also Refs.\cite{TanakaIch,introWD03}.}\label{fig:integration_contour}
\end{figure}

\emph{-- Derivation}. The proof strongly resembles the derivation of the infinite Matsubara sum for equal-time correlation functions\,\cite{quantelec,introWD03} with the zero frequency moment sum rule substituted by the two-sided Laplace transform expression for the ITCF, $F_{\mathrm{AA}^{\dagger}}(\tau)=\langle\hat{A}(-\imath\hbar\tau)\hat{A}^{\dagger}\rangle_0$,
\begin{equation}
F_{\mathrm{AA}^{\dagger}}(\tau)=\int_{-\infty}^{+\infty}S_{\mathrm{AA}^{\dagger}}(\omega)e^{-\hbar\omega\tau}d\omega\,.\label{ITCFdef}
\end{equation}
Thus, with the exception of the integration contour that deserves special attention, our exposition will mostly focus on the unique features of the present derivation. Substitution of the quantum FDT, utilization of the odd parity property $\Im\{\chi_{AA^{\dagger}}(-\omega)\}=-\Im\{\chi_{AA^{\dagger}}(\omega)\}$, conversion to hyperbolic algebra and use of the addition identity $\sinh(x)+\sinh(y)=2\sinh{[(x+y)/2]}\cosh{[(x-y)/2]}$, ultimately yield
\begin{align*}
F_{\mathrm{AA}^{\dagger}}(\tau)=-\frac{\hbar}{2\pi}\int_{-\infty}^{+\infty}\frac{\cosh{\left[\displaystyle\frac{\beta\hbar\omega}{2}-\hbar\omega\tau\right]}}{\sinh{\left(\displaystyle\frac{\beta\hbar\omega}{2}\right)}}\Im\{\chi_{AA^{\dagger}}(\omega)\}d\omega\,.
\end{align*}
We consider the contour integral $\oint_{\mathcal{C}}f(z)dz$ where $f(z)=\cosh{\left[(\beta\hbar{z}/2)-\hbar\tau{z}\right]}\mathrm{csch}{\left(\beta\hbar{z}/2\right)}\widetilde{\chi}_{AA^{\dagger}}(z)$ with $\widetilde{\chi}_{AA^{\dagger}}(z)$ the analytic continuation of the linear response function in the complex frequency domain $z=\omega+\imath\gamma$. The closed contour $\mathcal{C}$ should satisfy two restrictions: include the infinite simple poles of $\mathrm{csch}{\left(\beta\hbar{z}/2\right)}$ that are placed along the entire imaginary axis and exclude the branch cut of $\widetilde{\chi}_{AA^{\dagger}}(z)$ that lies infinitesimally below the real axis (arising from the merging of infinitely many poles of vanishing strength within the thermodynamic limit\,\cite{quantelec}). The chosen parametric $\mathcal{C}(\epsilon,\eta,R)$ contour has been illustrated in Fig.\ref{fig:integration_contour}. When traversed counterclockwise, it comprises a semi-circle $\Gamma_{\mathrm{u}}(R;0,\imath\eta)$ centered at $(0,\imath\eta)$ with radius $R$, a horizontal line segment $S_1$ at $+\imath\eta$ within $(-R,-\epsilon)$, a vertical line segment $S_2$ at $-\epsilon$ within $(+\imath\eta,-\imath\eta)$, a horizontal line segment $S_3$ at $-\imath\eta$ within $(-\epsilon,-R)$, a semi-circle $\Gamma_{\mathrm{l}}(R;0,-\imath\eta)$ centered at $(0,-\imath\eta)$ with radius $R$, a horizontal line segment $S_4$ at $-\imath\eta$ within $(R,\epsilon)$, a vertical line segment $S_5$ at $+\epsilon$ within  $(-\imath\eta,+\imath\eta)$, a horizontal line segment $S_6$ at $+\imath\eta$ within $(\epsilon,R)$. The desired contour is described by $\mathcal{C}(0,0,\infty)=\displaystyle\lim_{\eta\to0}\displaystyle\lim_{\epsilon\to0}\displaystyle\lim_{R\to\infty}\mathcal{C}(\epsilon,\eta,R)$, this iterated limit guarantees that the hyperbolic cosecant pole at $z=0$ is included within the integration contour, while the branch cut remains excluded. The contributions from the line segments $S_1$ and $S_6$, for which $z=\omega+\imath\eta$, yield a Cauchy principal value courtesy of the $\epsilon\to0$ limit:
\begin{align*}
\int_{S_1(0,0,\infty)}f(z)dz+\int_{S_6(0,0,\infty)}f(z)dz&=\displaystyle\lim_{\eta\to0}\mathcal{P}\int_{-\infty}^{+\infty}f(\omega+\imath\eta)d\omega\,.\nonumber
\end{align*}
The contributions from the line segments $S_3$ and $S_4$, for which $z=\omega-\imath\eta$, yield a Cauchy principal value courtesy of the $\epsilon\to0$ limit:
\begin{align*}
\int_{S_3(0,0,\infty)}f(z)dz+\int_{S_4(0,0,\infty)}f(z)dz&=-\displaystyle\lim_{\eta\to0}\mathcal{P}\int_{-\infty}^{+\infty}f(\omega-\imath\eta)d\omega\,.\nonumber
\end{align*}
The contributions from the line segments $S_2$ and $S_5$, for which $z=\mp\epsilon+\imath\gamma$, respectively, cancel out:
\begin{align*}
\int_{S_2(0,0)}f(z)dz+\int_{S_5(0,0)}f(z)dz&=0\,.\nonumber
\end{align*}
The contributions from the semi-circles $\Gamma_{\mathrm{u}}$, for which $z=\imath\eta+Re^{\imath\theta}$ with $\theta\in[0,\pi]$, and $\Gamma_{\mathrm{l}}$, for which $z=-\imath\eta+Re^{\imath\theta}$ with $\theta\in[-\pi,0]$, are also zero for $R\to\infty$ courtesy of the estimation lemma:
\begin{align*}
\int_{\Gamma_{\mathrm{u}}(\infty;0,\imath0)}f(z)dz=\int_{\Gamma_{\mathrm{l}}(\infty;0,-\imath0)}f(z)dz&=0\,.\nonumber
\end{align*}
Combining the above, substituting for the expression for the $\widetilde{\chi}_{AA^{\dagger}}(z)$ discontinuity jump across the real axis\,\cite{quantelec}, $\displaystyle\lim_{\eta\to0}\{\widetilde{\chi}_{AA^{\dagger}}(\omega+\imath\eta)-\widetilde{\chi}_{AA^{\dagger}}(\omega-\imath\eta)\}=2\imath\Im\{{\chi}_{AA^{\dagger}}(\omega)\}$, and utilizing the hyperbolic function expression for $F_{\mathrm{AA}^{\dagger}}(\tau)$, we acquire
\begin{equation}
\oint_{\mathcal{C}(0,0,\infty)}f(z)dz=-\frac{4\pi\imath}{\hbar}F_{AA^{\dagger}}(\tau)\,.\label{ITCFint1}
\end{equation}
We evaluate the contour integral at the LHS of Eq.(\ref{ITCFint1}) with the standard techniques of complex analysis. First, it is convenient to employ the argument subtraction identity $\cosh{(x-y)}=\cosh{(x)}\cosh{(y)}-\sinh{(x)}\sinh{(y)}$, which leads to the decomposition $f(z)=f_1(z)-f_2(z)$, with $f_1(z)=\cosh{\left(\hbar\tau{z}\right)}\coth{\left(\beta\hbar{z}/2\right)}\widetilde{\chi}_{AA^{\dagger}}(z)$ a meromorphic function within $\mathcal{C}(0,0,\infty)$ courtesy of the hyperbolic cotangent and with $f_2(z)=\sinh{\left(\hbar\tau{z}\right)}\widetilde{\chi}_{AA^{\dagger}}(z)$ an analytic function within $\mathcal{C}(0,0,\infty)$, since $\sinh{\left(\hbar\tau{z}\right)}$ is an entire function and $\widetilde{\chi}_{AA^{\dagger}}(z)$ is analytic within $\mathcal{C}(0,0,\infty)$. From the Cauchy-Goursat theorem, the latter leads to
\begin{align*}
\oint_{\mathcal{C}(0,0,\infty)}f_2(z)dz=0\,.
\end{align*}
From Cauchy's residue theorem, the former leads to
\begin{align*}
\oint_{\mathcal{C}(0,0,\infty)}f_1(z)dz=2\pi\imath\displaystyle\sum_{l}\mathrm{Res}\left\{f_1(z),z_l\right\}\,.
\end{align*}
The infinite simple poles $z_l$ of the $\coth{\left(\beta\hbar{z}/2\right)}$ factor lie on the imaginary frequency axis and their magnitude is given by the bosonic Matsubara frequencies $z_l=\imath\omega_l$ with $\omega_l=2\pi{l}/(\beta\hbar),\,\forall{l}\in\mathds{Z}$. In addition, when considering the pure imaginary nature of $\tau=(\imath/\hbar)t$, the infinite simple roots $z_m$ of the $\cosh{\left(\hbar\tau{z}\right)}$ factor lie on the real frequency axis. Furthermore, with the aid of the spectral representation, it can be proven that $\widetilde{\chi}_{AA^{\dagger}}(z)$ does not have any roots on the imaginary frequency axis\,\cite{quantelec}. Consequently, none of the infinite simple poles of the hyperbolic cotangent are removable. The respective residues are straightforwardly evaluated with the aid of L'Hopital's rule and basic hyperbolic function properties. This leads to
\begin{align*}
\oint_{\mathcal{C}(0,0,\infty)}f_1(z)dz=\frac{4\pi\imath}{\beta\hbar}\displaystyle\sum_{l=-\infty}^{\infty}\cosh{\left(\imath\hbar\omega_l\tau\right)}\widetilde{\chi}_{AA^{\dagger}}(\imath\omega_l)\,.
\end{align*}
After using the elementary property $\cosh{(\imath{x})}=\cos{(x)}$, the evenness of $\widetilde{\chi}_{AA^{\dagger}}(\imath\omega_l)$ with respect to the Matsubara order $l$ together with the oddness of $\sin{\left(\hbar\omega_l\tau\right)}$ with respect to the Matsubara order $l$ imply that the Matsubara series of $\sin{\left(\hbar\omega_l\tau\right)}\widetilde{\chi}_{AA^{\dagger}}(\imath\omega_l)$ is equal to zero and opens the way for the application of the Euler formula. Thus,
\begin{equation}
\oint_{\mathcal{C}(0,0,\infty)}f(z)dz=\frac{4\pi\imath}{\beta\hbar}\displaystyle\sum_{l=-\infty}^{\infty}\widetilde{\chi}_{AA^{\dagger}}(\imath\omega_l)e^{-\imath\hbar\omega_l\tau}\,.\label{ITCFint2}
\end{equation}
Combining Eqs.(\ref{ITCFint1},\ref{ITCFint2}) and solving for $F_{AA^{\dagger}}(\tau)$, we end up with the Fourier-Matsubara series expansion of the ITCF associated with an arbitrary operator $\hat{A}$
\begin{equation}
F_{AA^{\dagger}}(\tau)=-\frac{1}{\beta}\displaystyle\sum_{l=-\infty}^{\infty}\widetilde{\chi}_{AA^{\dagger}}(\imath\omega_l)e^{-\imath\hbar\omega_l\tau},\,\,\,\,\omega_l=\frac{2\pi{l}}{\beta\hbar}\,.\label{ITCFgeneral}
\end{equation}
It is apparent that this Fourier--Matsubara series expansion for the ITCFs, see Eq.(\ref{ITCFgeneral}), constitutes the generalization of the Matsubara series for the SSFs, see Eq.(\ref{SSF_Matsubara}), since $F_{AA^{\dagger}}(0)=S_{AA^{\dagger}}$. The connection to thermal quantum field theory is also evident, since our ITCF representation is reminiscent of Fourier expansions of the single-particle finite temperature Green's function (for bosons and fermions) in the Matsubara formalism\,\cite{greenMat1,greenMat2}.

\emph{-- Properties}. Apart from the reality property and the normalization property, the Fourier--Matsubara series expansion is also consistent with the imaginary--time symmetry property $F_{AA^{\dagger}}(\beta-\tau)=F_{AA^{\dagger}}(\tau)$ (which is a trivial consequence of the $2\pi$ periodicity of the cosine) as well as the imaginary--time version of the fluctuation--dissipation theorem\,\cite{introWD05,imagivFDT}. This is demonstrated by integrating the Fourier-cosine form of Eq.(\ref{ITCFgeneral}) within $\tau\in[0,\beta]$, singling out the static term and carrying out the trivial integrations which cancel all $\l\neq0$ contributions, thus yielding
\begin{equation}
\chi_{AA^{\dagger}}(0)=-\int_0^{\beta}F_{AA^{\dagger}}(\tau)d\tau\,.\label{ITCFinv0}
\end{equation}
Unsurprisingly, since the Fourier--Matsubara expansion essentially constitutes a complex Fourier series, the coefficients $\widetilde{\chi}_{AA^{\dagger}}(\imath\omega_l)$ can be expressed as integrals of the ITCF. The inversion formula is based on the orthonormality of complex exponentials and reads as
\begin{equation}
\widetilde{\chi}_{AA^{\dagger}}(\imath\omega_l)=-\int_0^{\beta}F_{AA^{\dagger}}(\tau)e^{\imath\hbar\omega_l\tau}d\tau\,.\label{ITCFinv1}
\end{equation}
Alternatively, the Fourier--Matsubara expansion can also be viewed as a cosine Fourier series, which leads to the equivalent inversion formula
\begin{equation}
\widetilde{\chi}_{AA^{\dagger}}(\imath\omega_l)=-2\int_0^{\beta/2}F_{AA^{\dagger}}(\tau)\cos{\left(\hbar\omega_l\tau\right)}d\tau\,,\label{ITCFinv2}
\end{equation}
where the imaginary--time symmetry property $F_{AA^{\dagger}}(\beta-\tau)=F_{AA^{\dagger}}(\tau)$ has also been employed. It is straightforward that the two inversion formulas constitute the arbitrary Matsubara order generalizations of the imaginary--time version of the fluctuation--dissipation theorem, compare Eqs.(\ref{ITCFinv1},\ref{ITCFinv2}) with Eq.(\ref{ITCFinv0}). Finally, it is also important to point out that the $n-$order frequency moments of the SF are directly connected with the $n-$order derivatives of the ITCF at $\tau=0$\,\cite{introWD04,introWD07}. The term-by-term differentiation of the cosine Fourier--Matsubara series expansion is not permitted and would erroneously suggest that all odd order frequency moments of the SF are identically zero. Conversely, it is evident that the Fourier--Matsubara series expansion cannot provide any information about the positive integer frequency moments of the SF.

\emph{-- Discussion}. The path from the time correlations associated with an arbitrary operator $\hat{A}$ to the time correlations associated with the density operator $\hat{\rho}(\boldsymbol{q})$, including spatial dispersion, is straightforward\,\cite{quantelec}. The substitutions
$F(\boldsymbol{q},\tau)\to(1/N)F_{\hat{\rho}(\boldsymbol{q})\hat{\rho}^{\dagger}(\boldsymbol{q})}(\tau)$ and $\chi(\boldsymbol{q},\omega)\to(1/V)\chi_{\hat{\rho}(\boldsymbol{q})\hat{\rho}^{\dagger}(\boldsymbol{q})}$ lead to the Fourier--Matsubara series expansion associated with the density operator $\hat{\rho}(\boldsymbol{q})$
\begin{equation}
F(\boldsymbol{q},\tau)=-\frac{1}{n\beta}\displaystyle\sum_{l=-\infty}^{\infty}\widetilde{\chi}(\boldsymbol{q},\imath\omega_l)e^{-\imath\hbar\omega_l\tau},\,\,\,\,\omega_l=\frac{2\pi{l}}{\beta\hbar}\,.\label{ITCFdensity}
\end{equation}

Let us first discuss the consequences of the Fourier-Matsubara series expansion for the finite temperature dielectric formalism\,\cite{IchiRepor,IchiRevMP,TanakaIch}. This framework combines the density version of the exact Eq.(\ref{SSF_Matsubara}),
\begin{equation}
S(\boldsymbol{q})=-\frac{1}{n\beta}\sum_{l=-\infty}^{+\infty}\widetilde{\chi}(\boldsymbol{q},\imath\omega_l)\,,\label{SSF_Matsubara_density}
\end{equation}
with the exact polarization potential approach expression for the density response function in terms of the ideal (Lindhard) density response $\chi_0(\boldsymbol{q},\omega)$ and the dynamic local field correction $G(\boldsymbol{q},\omega)$ (LFC)\,\cite{IchiRepor,IchiRevMP}
\begin{equation}
\chi(\boldsymbol{q},\omega)=\frac{\chi_0(\boldsymbol{q},\omega)}{1-U(\boldsymbol{q})\left[1-G(\boldsymbol{q},\omega)\right]\chi_0(\boldsymbol{q},\omega)}\,,\label{densityresponseDLFC}
\end{equation}
where $U(\boldsymbol{q})$ is the Fourier transformed pair interaction energy. The introduction of an approximation expression for the LFC as a functional of the SSF\,\cite{IchiRepor,IchiRevMP}
\begin{equation}
G(\boldsymbol{q},\omega)\equiv{G}[S](\boldsymbol{q},\omega)\,.\label{functionalclosure}
\end{equation}
then generates a closed set of equations to be solved for the SSF. The resulting non-linear functional equation is
\begin{equation}
S(\boldsymbol{q})=-\frac{1}{{n}\beta}\displaystyle\sum_{l=-\infty}^{\infty}\frac{\widetilde{\chi}_0(\boldsymbol{q},\imath\omega_l)}{1-U(\boldsymbol{q})\left[1-{G}[S](\boldsymbol{q},\imath\omega_l)\right]\widetilde{\chi}_0(\boldsymbol{q},\imath\omega_l)}\,.\nonumber
\end{equation}
Thus, without the Fourier--Matsubara series expansion of Eq.(\ref{ITCFdensity}), in order to compute the density--density ITCF, it would be necessary: to use the SSF for the calculation of the dynamic complex LFC in the frequency domain via Eq.(\ref{functionalclosure}), to employ the LFC to evaluate the dynamic complex density response function in the frequency domain via Eq.(\ref{densityresponseDLFC}), to use the imaginary part of the density response function to compute the DSF through the quantum FDT of Eq.(\ref{qFDT}), to apply the two-sided Laplace transform that comprises an involved frequency integration over the sharp quasi-particle peaks. On the other hand, by utilizing the Fourier--Matsubara series expansion of Eq.(\ref{ITCFdensity}), the density-density ITCF is directly computed without any extra steps and at no computational cost, since the converged $\widetilde{\chi}(\boldsymbol{q},\imath\omega_l)$ should be readily available from the SSF numerical solution.

Let us also discuss the consequences of the inversion formula of the Fourier-Matsubara series expansion for the post-processing of QMC simulations. The translation of the inversion formula from the general linear response theory to density response theory yields
\begin{equation}
\widetilde{\chi}(\boldsymbol{q},\imath\omega_l)=-n\int_0^{\beta}F(\boldsymbol{q},\tau)e^{\imath\hbar\omega_l\tau}d\tau\,.\label{ITCFinv1density}
\end{equation}
Consequently, QMC data for the density-density ITCF can be employed to evaluate the dynamic density response function at the imaginary Matsubara frequencies. Courtesy of the constitutive relation of the polarization approach, see Eq.(\ref{densityresponseDLFC}), this implies that QMC simulations provide direct access to the quasi-exact dynamic LFC at imaginary Matsubara frequencies. Such information would not only constitute a stringent benchmarking test for theoretical approaches, but would also constitute the basis for analytical representations of the dynamic LFC. The importance of such a feat cannot be understated; indeed, it is worth pointing out that analytical representations are only available for the frequency averaged LFC and only in the WDM phase diagram subregion of the UEG\,\cite{ESAschem3}. Such future parametrizations of the dynamic LFC based on quasi-exact QMC simulations will open up new ways for the construction of advanced, non-local, and consistently thermal exchange--correlation functionals for density functional theory (DFT) simulations based on the adiabatic connection formula and the fluctuation--dissipation theorem~\cite{PribramJo}. Moreover, such a parametrization would allow to use DFT---an effective single-electron theory---for the estimation of electron--electron correlation functions, thereby adding a new dimension to one of the most successful simulation techniques in quantum chemistry and related fields. Finally, we mention the intriguing possibility to extend the present linear response result to the non-linear case, given the known connection between dynamic non-linear responses and higher-order imaginary--time correlation functions~\cite{Nonlinea1,Nonlinea2,Nonlinea3}.

\emph{-- Acknowledgments}. This work was partially supported by the Center for Advanced Systems Understanding (CASUS) which is financed by Germany's Federal Ministry of Education and Research (BMBF) and the Saxon state government out of the State budget approved by the Saxon State Parliament (CASUS Open Project: Guiding dielectric theories with ab initio quantum Monte Carlo simulations: from the strongly coupled electron liquid to warm dense matter). This work has also received funding from the European Research Council (ERC) under the European Union’s Horizon 2022 research and innovation programme (Grant agreement No.101076233, "PREXTREME"). Views and opinions expressed are those of the authors and do not necessarily reflect those of the European Union or the European Research Council Executive Agency. Neither the European Union nor the granting authority can be held responsible for them.

\emph{-- Data availability}. Data sharing is not applicable to this article as no new data were created or analyzed in this study.


\begin{thebibliography}{200}
\bibitem{introQMC1} D. M. Ceperley, Path integrals in the theory of condensed helium, \emph{Rev. Mod. Phys.} {\bf 67}, 279 (1995).
\bibitem{introQMC2} M. F. Herman, E. J. Bruskin, and B. J. Berne, On path integral Monte Carlo simulations, \emph{J. Chem. Phys.} {\bf 76}, 5150 (1982).
\bibitem{introQMC3} T. Dornheim, S. Groth, A. V. Filinov, and M. Bonitz, Path integral Monte Carlo simulation of degenerate electrons: Permutation-cycle properties, \emph{J. Chem. Phys.} {\bf 151}, 014108 (2019).
\bibitem{introAC01} M. Jarrell and J. E. Gubernatis, Bayesian inference and the analytic continuation of imaginary-time quantum Monte Carlo data, \emph{Phys. Rep.} {\bf 269}, 133 (1996).
\bibitem{introAC02} H. Shao and A. W. Sandvik, Progress on stochastic analytic continuation of quantum Monte Carlo data, \emph{Phys. Rep.} {\bf 1003}, 1 (2023).
\bibitem{introAC03} R. N. Silver, D. S. Sivia, and J. E. Gubernatis, Maximum-entropy method for analytic continuation of quantum Monte Carlo data, \emph{Phys. Rev. B} {\bf 41}, 2380 (1990).
\bibitem{introAC04} S. Fuchs, T. Pruschke, and M. Jarrell, Analytic continuation of quantum Monte Carlo data by stochastic analytical inference, \emph{Phys. Rev. E} {\bf 81}, 056701 (2010).
\bibitem{introAC05} Y. Kora and M. Boninsegni, Dynamic structure factor of superfluid 4He from quantum Monte Carlo: Maximum entropy revisited, \emph{Phys. Rev. B} {\bf 98}, 134509 (2018).
\bibitem{introAC06} A. W. Sandvik, Constrained sampling method for analytic continuation, \emph{Phys. Rev. E} {\bf 94}, 063308 (2016).
\bibitem{introAC07} G. Bertaina, Davide Emilio Galli, and E. Vitali, Statistical and computational intelligence approach to analytic continuation in quantum Monte Carlo, \emph{Adv. Phys.: X} {\bf 2}, 302 (2017).
\bibitem{introAC08} I. Krivenko and M. Harland, TRIQS/SOM: Implementation of the stochastic optimization method for analytic continuation, \emph{Comput. Phys. Commun.} {\bf 239}, 166 (2019).
\bibitem{introAC09} J. Otsuki, M. Ohzeki, H. Shinaoka, and K. Yoshimi, Sparse modeling approach to analytical continuation of imaginary-time quantum Monte Carlo data, \emph{Phys. Rev. E} {\bf 95}, 061302(R) (2017).
\bibitem{introAC10} J. Fei, Chia-Nan Yeh, and E. Gull, Nevanlinna Analytical Continuation, \emph{Phys. Rev. Lett.} {\bf 126}, 056402 (2021).
\bibitem{introUEG1} T. Dornheim, S. Groth, J. Vorberger, and M. Bonitz, \emph{Ab initio} path integral Monte Carlo results for the dynamic structure factor of correlated electrons: From the electron liquid to warm dense matter, \emph{Phys. Rev. Lett.} {\bf 121}, 255001 (2018).
\bibitem{introUEG2} S. Groth, T. Dornheim, and J. Vorberger, \emph{Ab initio} path integral Monte Carlo approach to the static and dynamic density response of the uniform electron gas, \emph{Phys. Rev. B} {\bf 99}, 235122 (2019).
\bibitem{introWD01} F. Graziani, M. P. Desjarlais, R. Redmer, and S. B. Trickey, \emph{Frontiers and Challenges in Warm Dense Matter}, (Springer International, Switzerland, 2014).
\bibitem{introWD02} M. Bonitz, T. Dornheim, Zh. A. Moldabekov, S. Zhang \emph{et al.}, \emph{Ab initio} simulation of warm dense matter, \emph{Phys. Plasmas} {\bf 27}, 042710 (2020).
\bibitem{introWD03} T. Dornheim, S. Groth, and M. Bonitz, The uniform electron gas at warm dense matter conditions, \emph{Phys. Rep.} {\bf 744}, 1 (2018).
\bibitem{introWD04} T. Dornheim, Zh. A. Moldabekov, K. Ramakrishna, P. Tolias \emph{et al.}, Electronic density response of warm dense matter, \emph{Phys. Plasmas} {\bf 30}, 032705 (2023).
\bibitem{introWD05} T. Dornheim, Zh. Moldabekov, P. Tolias, M. B{\"o}hme, and Jan Vorberger, Physical insights from imaginary-time density-density correlation functions, \emph{Matter Radiat. Extremes} {\bf 8}, 056601 (2023).
\bibitem{introWD06} T. Dornheim, J. Vorberger, Z. A. Moldabekov, and M. B{\"o}hme, Analysing the dynamic structure of warm dense matter in the imaginary-time domain: Theoretical models and simulations, \emph{Philos. Trans. R. Soc. A} {\bf 381}, 20220217 (2023).
\bibitem{introWD07} T. Dornheim, D. C. Wicaksono, J. E. Suarez-Cardona, P. Tolias \emph{et al.}, Extraction of the frequency moments of spectral densities from imaginary-time correlation function data, \emph{Phys. Rev. B} {\bf 107}, 155148 (2023).
\bibitem{introWD08} T. Dornheim, M. B{\"o}hme, D. Kraus, T. D{\"o}ppner \emph{et al.}, Accurate temperature diagnostics for matter under extreme conditions, \emph{Nat. Commun.} {\bf 13}, 7911 (2022).
\bibitem{introWD09} T. Dornheim, M. P. B{\"o}hme, D. Chapman, D. Kraus \emph{et al.}, Imaginary-time correlation function thermometry: A new, high-accuracy and model-free temperature analysis technique for x-ray Thomson scattering data, \emph{Phys. Plasmas} {\bf 30}, 042707 (2023).
\bibitem{introWD10} T. Dornheim, T. D{\"o}ppner, A. D. Baczewski, P. Tolias \emph{et al.}, X-ray thomson scattering absolute intensity from the f-sum rule in the imaginary-time domain, arXiv:2305.15305 [physics.plasm-ph] (2024).
\bibitem{introWD11} S. H. Glenzer and R. Redmer, X-ray Thomson scattering in high energy density plasmas, \emph{Rev. Mod. Phys.} {\bf 81}, 1625 (2009).
\bibitem{introWD12} J. Sheffield, D. Froula, S. H. Glenzer, and N. C. Luhmann, Plasma Scattering of Electromagnetic Radiation: Theory and Measurement Techniques (Elsevier Science, 2010).
\bibitem{introWD13} J. Chihara, Difference in x-ray scattering between metallic and non-metallic liquids due to conduction electrons, \emph{J. Phys. F: Met. Phys.} {\bf 17}, 295 (1987).
\bibitem{introWD14} G. Gregori, S. H. Glenzer, W. Rozmus, R. W. Lee, and O. L. Landen, Theoretical model of x-ray scattering as a dense matter probe, \emph{Phys. Rev. E} {\bf 67}, 026412 (2003).
\bibitem{introWD1f} M. B{\"o}hme, L. Fletcher, T. D{\"o}ppner, D. Kraus \emph{et al.}, Evidence of free-bound transitions in warm dense matter and their impact on equation-of-state measurements, arXiv:2306.17653 [physics.plasm-ph] (2023).
\bibitem{introWD15} T. Dornheim, T. D{\"o}ppner, P. Tolias, M. B{\"o}hme \emph{et al.}, Unraveling electronic correlations in warm dense quantum plasmas, arXiv:2402.19113 [physics.plasm-ph] (2024).
\bibitem{introWD16} T. D{\"o}ppner, M. Bethkenhagen, D. Kraus, P. Neumayer \emph{et al.}, Observing the onset of pressure-driven k-shell delocalization, \emph{Nature} {\bf 618}, 270275 (2023).
\bibitem{NoziPines} P. Nozieres and D. Pines, A dielectric formulation of the many body problem: Application to the free electron gas, \emph{Il Nuovo Cimento}  {\bf 9}, 470 (1958).
\bibitem{SingwTosi} K. S. Singwi and M. P. Tosi, Correlations in electron liquids, \emph{Solid State Phys.} {\bf 36}, 177 (1981).
\bibitem{IchiRepor} S. Ichimaru, H. Iyetomi, and S. Tanaka, Statistical physics of dense plasmas: Thermodynamics, transport coefficients and dynamic correlations, \emph{Phys. Rep.} {\bf 149}, 91 (1987).
\bibitem{IchiRevMP} S. Ichimaru, Nuclear fusion in dense plasmas, \emph{Rev. Mod. Phys.} {\bf 65}, 255 (1993).
\bibitem{quantelec} G. Giuliani and G. Vignale, \emph{Quantum theory of the electron liquid} (Cambridge University Press, 2008).
\bibitem{TanakaIch} S. Tanaka and S. Ichimaru, Thermodynamics and correlational properties of finite-temperature electron liquids in the Singwi-Tosi-Land-Sj{\"o}lander approximation, \emph{J. Phys. Soc. Jpn.} {\bf 55}, 2278 (1986).
\bibitem{qSTLSsche} H. K. Schweng and H. M. B\"ohm, Finite-temperature electron correlations in the framework of a dynamic local-field correction, \emph{Phys. Rev. B} {\bf 48}, 2037 (1993).
\bibitem{QMCresul1} T. Dornheim, S. Groth, T. Sjostrom, F. D. Malone \emph{et al.}, \emph{Ab initio} quantum Monte Carlo simulation of the warm dense electron gas in the thermodynamic limit, \emph{Phys. Rev. Lett.} {\bf 117}, 156403 (2016).
\bibitem{QMCresul2} S. Groth, T. Dornheim, T. Sjostrom, F. D. Malone \emph{et al.}, \emph{Ab initio} exchange-correlation free energy of the uniform electron gas at warm dense matter conditions, \emph{Phys. Rev. Lett.} {\bf 119}, 135001 (2017).
\bibitem{ESAschem1} T. Dornheim, J. Vorberger, S. Groth, N. Hoffmann \emph{et al.}, The static local field correction of the warm dense electron gas: An \emph{ab initio} path integral Monte Carlo study and machine learning representation, \emph{J. Chem. Phys} {\bf 151}, 194104 (2019).
\bibitem{ESAschem2} T. Dornheim, A. Cangi, K. Ramakrishna, M. B\"ohme \emph{et al.}, Effective static approximation: A fast and reliable tool for warm-dense matter theory, \emph{Phys. Rev. Lett.} {\bf 125}, 235001 (2020).
\bibitem{ESAschem3} T. Dornheim, Zh. A. Moldabekov, and P. Tolias, Analytical representation of the local field correction of the uniform electron gas within the effective static approximation, \emph{Phys. Rev. B} {\bf 103}, 165102 (2021).
\bibitem{VSscheme1} T. Sjostrom and J. Dufty, Uniform electron gas at finite temperatures, \emph{Phys. Rev. B} {\bf 88}, 115123 (2013).
\bibitem{VSscheme2} P. Tolias, F. Lucco Castello, F. Kalkavouras, and T. Dornheim, Revisiting the Vashishta-Singwi dielectric scheme for the warm dense uniform electron fluid, \emph{Phys. Rev. B} {\bf 109}, 125134 (2024).
\bibitem{HNCschem1} S. Tanaka, Correlational and thermodynamic properties of finite-temperature electron liquids in the hypernetted-chain approximation, \emph{J. Chem. Phys.} {\bf 145}, 214104 (2016).
\bibitem{HNCschem2} T. Dornheim, T. Sjostrom, S. Tanaka, and J. Vorberger, Strongly coupled electron liquid: \emph{Ab initio} path integral Monte Carlo simulations and dielectric theories, \emph{Phys. Rev. B} {\bf 101}, 045129 (2020).
\bibitem{IETschem1} P. Tolias, F. Lucco Castello, and T. Dornheim, Integral equation theory based dielectric scheme for strongly coupled electron liquids, \emph{J. Chem. Phys.} {\bf 155}, 134115 (2021).
\bibitem{IETschem2} F. Lucco Castello, P. Tolias, and T. Dornheim, Classical bridge functions in classical and quantum plasma liquids, \emph{EPL} {\bf 138}, 44003 (2022).
\bibitem{qIETschem} P. Tolias, F. Lucco Castello, and T. Dornheim, Quantum version of the integral equation theory-based dielectric scheme for strongly coupled electron liquids, \emph{J. Chem. Phys.} {\bf 158}, 141102 (2023).
\bibitem{greenMat1} G. D. Mahan, \emph{Many-Particle Physics} (Kluwer, New York, 2000).
\bibitem{greenMat2} A. L. Fetter, and J. D. Walecka, \emph{Quantum Theory of Many-Particle Systems} (Dover, New York, 2003).
\bibitem{imagivFDT} C. Bowen, G. Sugiyama, and B. J. Alder, Static dielectric response of the electron gas, \emph{Phys. Rev. B} {\bf 50}, 14838 (1994).
\bibitem{PribramJo} A. Pribram-Jones, P. Grabowski, and K. Burke, Thermal Density Functional Theory: Time-Dependent Linear Response and Approximate Functionals from the Fluctuation-Dissipation Theorem, \emph{Phys. Rev. Lett.} {\bf 116}, 233001 (2016)
\bibitem{Nonlinea1} T. Dornheim, Z. Moldabekov, and J. Vorberger, Nonlinear density response from imaginary-time correlation functions: Ab initio path integral Monte Carlo simulations of the warm dense electron gas, \emph{J. Chem. Phys.} {\bf 155}, 054110 (2021).
\bibitem{Nonlinea2} T. Dornheim, J. Vorberger, and Z. Moldabekov, Nonlinear Density Response and Higher Order Correlation Functions in Warm Dense Matter, \emph{J. Phys. Soc. Jpn.} {\bf 90}, 104002 (2021).
\bibitem{Nonlinea3} P. Tolias, T. Dornheim, Z. Moldabekov, and J. Vorberger, Unravelling the nonlinear ideal density response of many-body systems, \emph{EPL} {\bf 142} 44001 (2023).
\end{thebibliography}
\end{document}